\documentclass[aps,pre,twocolumn,english,superscriptaddress,showpacs,floatfix,groupedaddress]{revtex4}
\usepackage{graphicx,epsfig}
\usepackage{amssymb}

\begin{document}

\title{Spectral properties and pattern selection in
fractal growth networks}
\author{K. Tucci}
  \email{kay@ula.ve}
\affiliation{
  Max-Planck-Institut f\"ur Physik Komplexer Systeme, 
      N\"othnizer Strasse 38,
      01187 Dresden, Germany} 
\affiliation{
  SUMA-CeSiMo, Universidad de Los Andes, 
      M\'erida 5251, Venezuela} 
\author{M. G. Cosenza}
\affiliation{
  Centro de Astrof\'{\i}sica Te\'orica, Facultad de
           Ciencias, Universidad de Los Andes,
           Apartado Postal 26 La Hechicera, M\'erida 5251, Venezuela
}

\begin{abstract}
A model for the generation of fractal growth networks in Euclidean
spaces of arbitrary dimension is presented. These networks are
considered as the spatial support of reaction-diffusion and
pattern formation processes. The local dynamics at the nodes of a
fractal growth network is given by a nonlinear map, giving raise
to a coupled map system. The coupling is described by a matrix
whose eigenvectors constitute a basis on which spatial patterns on
fractal growth networks can be expressed by linear combination.
The spectrum of eigenvalues the coupling matrix exhibits a
nonuniform distribution that is reflected in the presence of gaps
or niches in the boundaries of stability of the synchronized
states on the space of parameters of the system. These gaps allow
for the selection of specific spatial patterns by appropriately
varying the parameters of the system.
\end{abstract}

\maketitle

\section{Introduction}
Pattern formation processes in regimes far from equilibrium often
take place on media that are nonuniform at some length scales. The
nonuniformity may be due to the intrinsic heterogeneous nature of
the substratum, typical of pattern formation in biological
contexts, or it may arise from random imperfections or
fluctuations in the medium. Such heterogeneities can have
significant effects on the form of spatial patterns, for example,
they can induce reberverators in excitable media and defects can
serve as nucleation sites for domain growth processes. Recently,
there has been much interest in the study of spatiotemporal
dynamical processes on nonuniform or complex networks. In this
context, coupled map lattices \cite{Kaneko} have provided fruitful
and computationally efficient models for the investigation of a
variety of dynamical processes in spatially distributed systems.
In particular, the discrete-space character of coupled map systems
makes them specially appropriate for the investigation of
spatiotemporal dynamics on nonuniform networks that can represent
models of heterogeneous media. Phenomena such as pattern
formation, spatiotemporal intermittency, nontrivial collective
behavior, synchronization, phase-ordering, etc., have been
extensively studied in coupled map systems defined on fractal
lattices \cite{CK1,TCA}, hierarchical structures \cite{Gade},
trees \cite{Tree}, random graphs \cite{Vol}, small-world networks
\cite{Kay}, and scale-free networks \cite{Amri}.

An especially interesting class of nonuniform geometries comprises
fractal growth networks \cite{V1983} whose branching structure,
self-similar scaling features and lack of translation symmetry can
give rise to several distinct characteristics in both their
dynamical and spatial properties. Fractal growth structures appear
in nonequilibrium growth processes which are common in many areas.
Examples of such phenomena include diffusion-limited aggregation,
dendritic solidification, bacterial growth, viscous fingering,
capillarity, electrodeposition, Laplacian growth problems, etc.
\cite{Vicsek}.

In this article we consider discrete reaction-diffusion processes
and pattern formation taking place on fractal growth networks. The
spatiotemporal dynamics corresponds to a coupled map system
defined on the geometrical support of a fractal growth network.
Although many growth structures found in nature have random
features, here we study the case of simple, deterministic fractal
growth networks. We focus on the changes occurred in
spatiotemporal patterns as a result of the fractal-growth
connectivity that describes the interactions in the system. In
Sec.~II, a model for the construction of fractal growth networks
in any Euclidean space and a general notation for their treatment
are introduced. In Sec.~III, the coupled map models defined on
fractal growth networks are presented. The diffusion coupling
among neighboring sites on a network is described by a matrix. The
spectrum of eigenvalues and eigenvectors of the coupling matrix is
analyzed in Sec.~IV. The eigenvectors constitute a complete basis
on which spatial patterns can be expressed by a linear
combination. Section~V contains a study of the stability of
spatially uniform, periodic patterns on a fractal growth network
for a local dynamics given by the logistic map. Distinct features,
emerging as a consequence of the fractal nature of the networks,
allow for the selection of specific spatial patterns as the
parameters of the system are changed. Conclusions are presented in
Sec.~V.

\section{Model for fractal growth networks}
Fractal growth networks can be generated in any Euclidean space of
dimension $d$ by generalizing the fractal growth model of Vicsek
\cite{V1983}. Starting from a $d$-dimensional hypercube, it is
divided into $3^d$ equal hypercubes. Of these, only the central
hypercube plus the $2^d$ hypercubes connected to its vertices are
kept. This process is repeated for each of the $2^d+1$ resulting
hypercubes. At a level of construction $L$, the network consists
of $N=(2^d+1)^L$ hypercubical cells or nodes whose coordinates can
be specified by a sequence of symbols
$(\alpha_1\alpha_2\dots\alpha_L)$, where $\alpha_i$ can take any
value in a collection of $2^d+1$ different digits forming an
enumeration system in base $(2^d+1)$, which we denote by
$\{\epsilon_1,\epsilon_2\ldots, \epsilon_{2^{d+1}+1}\}$. For
example, for $d=2$, we may chose $\alpha_i \in \{0,1,2,3,4\}$.

To fix the labels, we start at level $L=1$ where the labels of the
$2^d + 1$ cells in the network have the form $(\alpha_1)$. The
label $(0)$ is assigned to the central cell, and the others $2^d$
cells are labeled by $(\alpha_i) \neq (0)$ in such a way that the
labels $(\alpha_i)$ and $(\alpha_j)$ of the two cells located at
the opposite ends of each diagonal passing through the central
cell satisfy $\mbox{mod}_{2^d+1}(\alpha_i+\alpha_j)=0$, where the
addition $\alpha_i+\alpha_j$ is defined modulo $(2^d+1)$.
Figures~1(a) and 1(b) show this notation for the fractal growth
networks embedded in Euclidean spaces of dimension $d=2$ and
$d=3$, respectively.
\begin{figure}[ht]
\centerline{\hbox{
\psfig{file=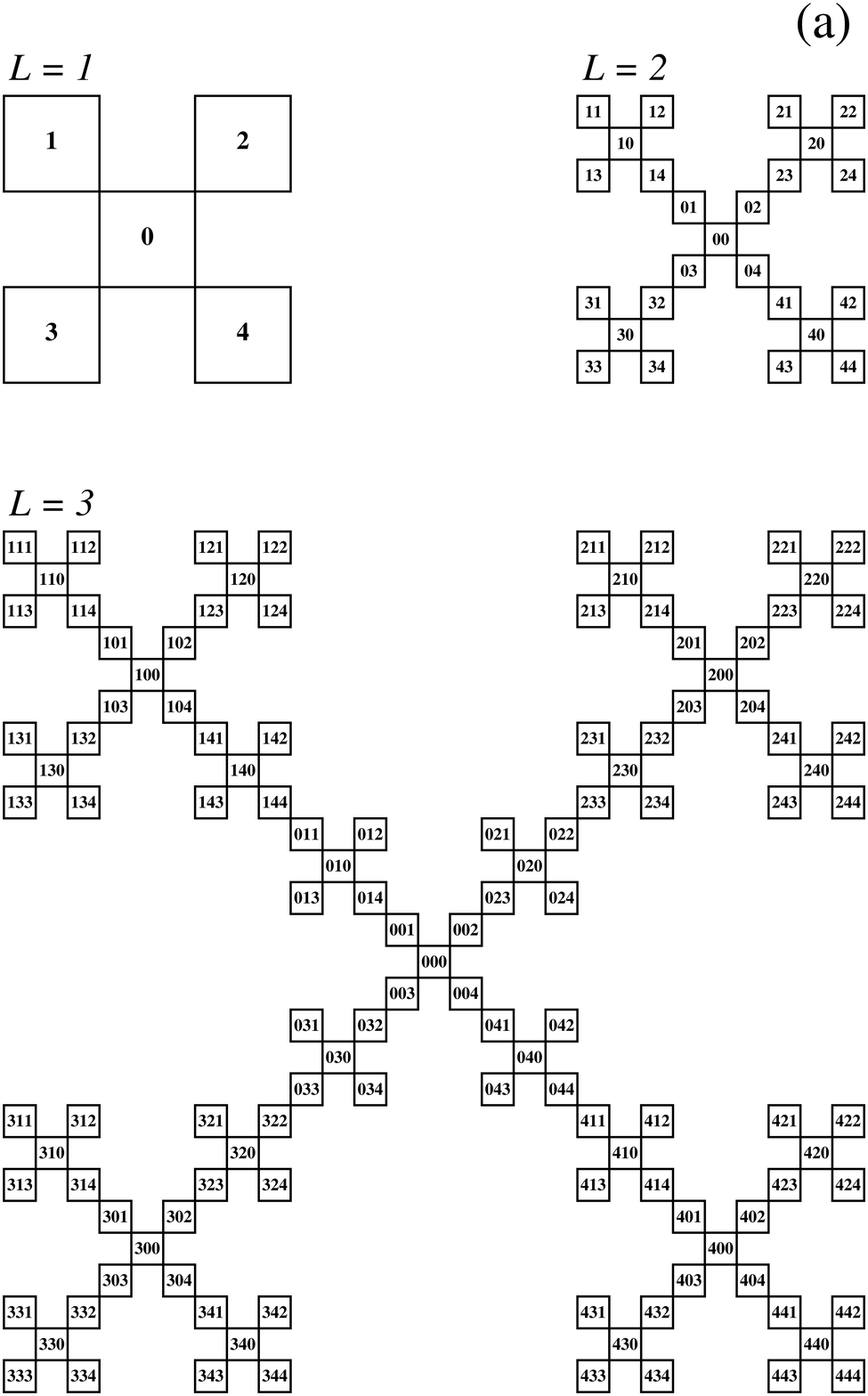,width=0.50\linewidth,clip=,angle=0}
}}\vspace{10mm}  
\centerline{\hbox{
\psfig{file=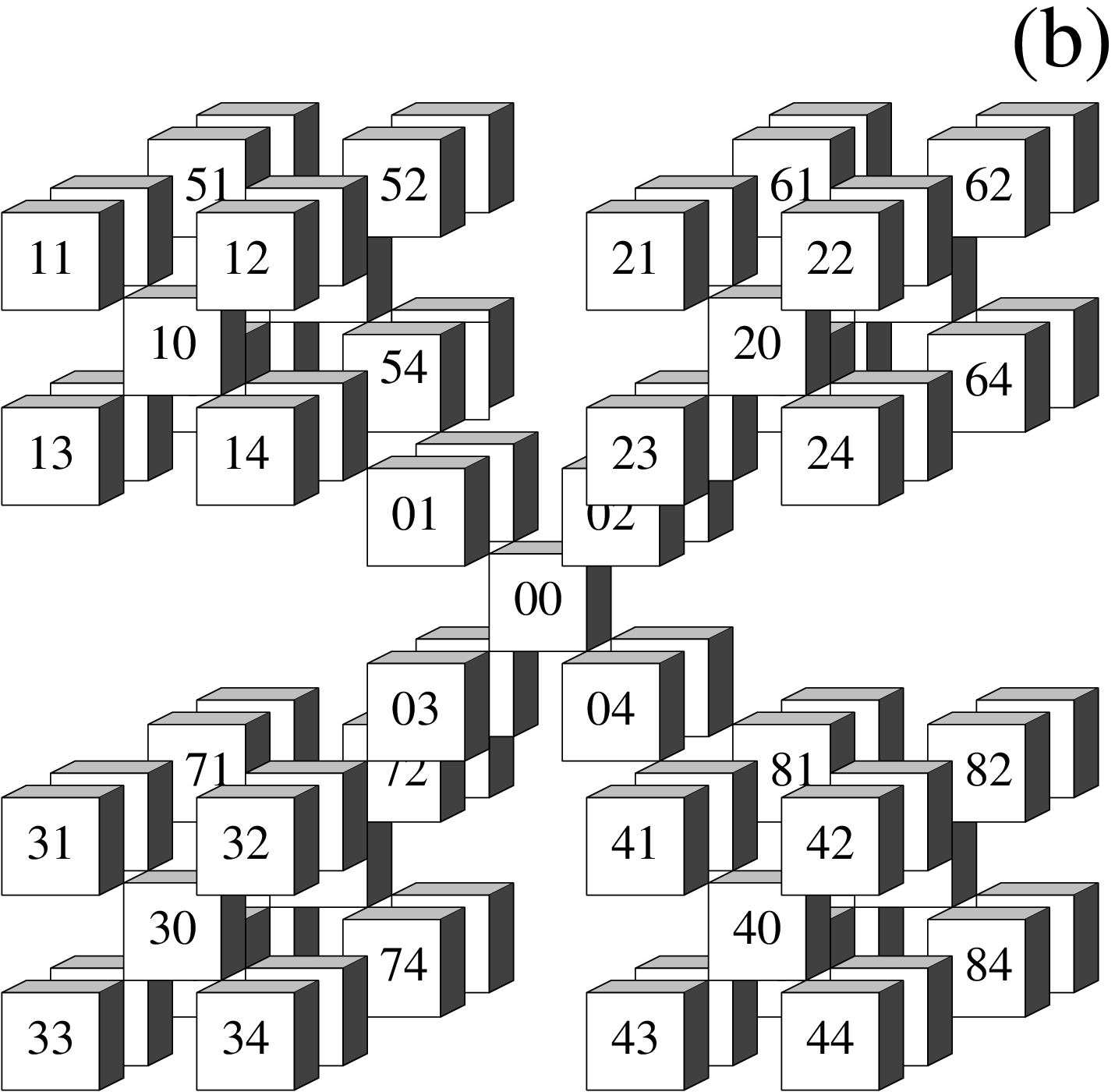,width=0.40\linewidth,clip=,angle=0}
}}
\caption{(a) Fractal growth network embedded in Euclidean space of
dimension $d=2$ at levels of construction $L=1$, $L=2$ and $L=3$,
showing labels on the cells.  (b) Fractal growth network embedded
in dimension $d=3$ at level of construction $L=2$, showing labels
on the cells.}
\end{figure}

When the fractal network grows from level of construction $L$ to
level $(L+1)$, each cell $(\alpha_1\alpha_2\dots\alpha_L)$ is
subdivided into $2^d+1$ cells, scaled down by a longitudinal
factor of $3$, and which are now labeled by
$(\alpha_1\alpha_2\dots\alpha_L\alpha_{L+1})$, where the first $L$
symbols of the sequence, $\alpha_1\alpha_2\dots\alpha_L$, are the
same as in the parent cell, and to distinguish the $(2^d+1)$
daughter cells, the last symbol $\alpha_{L+1}$ is assigned
similarly to the labels $(\alpha_1)$ at the construction level
$L=1$. From the construction process, it follows that the fractal
dimension of the network is $d_f=\ln(2^d+1)/\ln 3$.

The set of cells connected to the cell labeled by
$(\alpha_1\alpha_2\dots\alpha_L)$ at a level of construction $L$
is defined as the neighborhood (nearest neighbors) of this cell,
and will be denoted by ${\mathcal
N}(\alpha_1\alpha_2\dots\alpha_L)$. Three types of cells can be
identified in a fractal growth network embedded in a
$d$-dimensional Euclidean space: (i) centers, that are connected
to $2^d+1$ other cells; (ii) joints, that are connected to $2$
cells; and (iii) edges, connected to just one cell. A sequence
$(\alpha_1\alpha_2\dots\alpha_L)$ can be written as
$(\alpha_1\dots\alpha_{L-s}\alpha^s_{L-s+1})$ for some $s \in
\{1,2,\ldots,L-1\}$,  and where $\alpha_i^s$ means the sequence of
$s$ symbols $\alpha_i$. A center cell is labeled by a sequence
$(\alpha_1\alpha_2\dots\alpha_{L-s}\alpha^{s-1}_{L-s+1}0)$. Its
neighborhood set consists of $2^d$ cells, labeled by the sequences
$(\alpha_1\alpha_2\dots\alpha^{s-1}_{L-s+1}1)$,
$(\alpha_1\alpha_2\dots\alpha_{L-s}\alpha^{s-1}_{L-s+1}2)$,
$\ldots$, and
$(\alpha_1\alpha_2\dots\alpha_{L-s}\alpha^{s-1}_{L-s+1}2^d)$,
respectively. A joint cell labeled by
$(\alpha_1\dots\alpha_{L-s}\alpha^s_{L-s+1})$, with
$\alpha_{L-s+1} \neq 0$, has a neighborhood set with $2$ elements.
One is the center cell labeled by
$(\alpha_1\dots\alpha_{L-s}\alpha^{s-1}_{L-s+1}0)$. The other
neighbor is another joint cell labeled by
$(\alpha_1\dots\alpha_{L-s-1}\alpha_{L-s+1}\beta^s)$, if
$\alpha_{L-s} = 0$; or by $(\alpha_1\dots\alpha_{L-s-1}0\beta^s)$,
if $\alpha_{L-s} \ne 0$; where $\beta$ is an allowed symbol that
satisfies $\mbox{mod}_{2^d+1}(\alpha_{L-s+1}+\beta)=0$. This
procedure identifies any two joint cells as reciprocal neighbors.
On the other hand, an edge cell labeled by the sequence
$(\alpha_1\dots\alpha_{L-s}\alpha^s_{L-s+1})$ can not be assigned
a reciprocal neighbor from the above procedure; thus this edge
possesses only one neighbor, the center cell labeled by
$(\alpha_1\dots\alpha_{L-s}\alpha^{s-1}_{L-s+1}0)$.

Because the symbols $\alpha_i$ belong to an enumeration system in
base $(2^d+1)$, a cell in the network labeled by the sequence
$(\alpha_1 \alpha_2\dots\alpha_L)$ can be univocally associated to
an integer index $i=0,1,2,\ldots,(2^d+1)^L-1$, by the rule
\begin{equation}
(\alpha_1 \alpha_2\dots\alpha_L) \leftrightarrow i = \sum_{j=1}^L
\alpha_j (2^d+1)^{L-j} \, .
 \label{eq:rule}
\end{equation}

As an illustration, consider the fractal growth network embedded
in an Euclidean space of dimension $d=2$, as in Fig.~1(a). At the
level of construction $L=3$, the cell labeled by $(233)=(23^2)$ is
a joint cell and has index $i=68$, according to the rule
Eq.~(\ref{eq:rule}). Its two neighbors, labeled by $(230)$ (a
center cell) and $(022=(02^2)$ (another joint cell), are assigned
indexes $i=65$ and $i=12$, respectively. Similarly, the cell
$(422)=(42^2)$ is an edge with associated index $i=112$; its only
neighbor is the center cell $(420)$, corresponding to $i=110$.

\section{Coupled map lattice model for pattern formation}
A network might be considered as the spatial support of a dynamical
spatiotemporal processes with either discrete or continuous time.
Here we consider reaction-diffusion and pattern formation
phenomena on fractal growth networks. By associating a nonlinear
function to each cell of a given fractal network and coupling
these functions through nearest-neighbor diffusion interaction, we
define a coupled map lattice that describes a reaction-diffusion
dynamics as follows
\begin{equation}\label{eq:CML-snowflake}
 x_{t+1}(i) = f \left( x_t(i) \right) +
             \gamma\sum_{j \in {\mathcal N}(i)}
               \left( x_t(j) - x_t(i) \right) ,
\end{equation}
where $x_t(i)$ represents the state of the cell having index $i$,
assigned by Eq.~(\ref{eq:rule}), at discrete time $t$; $f(x_t(i))$
is a nonlinear function specifying the local dynamics; ${\mathcal
N}(i)$ is the neighborhood set of the cell with index $i$; and
$\gamma$ is a parameter expressing the coupling strength among
neighboring cells and it plays the role of a homogeneous diffusion
constant. The form of the coupling term in Eq.~(\ref{eq:rule}) is
usually called backward diffusive coupling and corresponds to a
discrete version of the Laplacian in reaction-diffusion equations.
This coupled map model can be generalized to include other
coupling schemes, nonuniform coupling, or continuous-time local
dynamics.

Equations~(\ref{eq:CML-snowflake}) can be written in a vector form
as
\begin{equation} \label{eq:vector}
{\bf x}_{t+1} = {\bf f}({\bf x}_t) + \gamma{\bf M}{\bf x}_t \, .
\end{equation}
The state vector ${\bf x}_t$ has $N$ components $[{\bf
x}_t]_i=x_t(i)$, $i=0,\dots,N-1$, corresponding to the states
$x_t(i)=x_t(\alpha_1\alpha_2\dots\alpha_L)$ of the cells on the
network. The $N \times N$ real, symmetric matrix ${\bf M}$
expresses the coupling among the $N$ components $x_t(i)$. For a
fractal growth network embedded in an Euclidean space of dimension
$d$ at the level of construction $L$, the components of the
corresponding matrix ${\bf M}$, denoted by $M(i,j)$ $(i,j =
0,1,\dots, N~-~1),$ are
\begin{equation}
M(i,j) = M(j,i) = \cases{
    1  ,                             & if $j \in {\mathcal N}(i)$  \cr
    -\left|{\mathcal N}(i)\right|,   & if $i=j$  \cr
    0    ,                           & elsewhere,}
\end{equation}
where $|{\mathcal N}(i)|$ is the cardinality of the neighborhood
set ${\mathcal N}(i)={\mathcal N}(\alpha_1\alpha_2\dots\alpha_L)$.
The matrix ${\bf M}$ plays the role of the spatially discrete
diffusion operator on these networks, similar to the Laplacian in
a spatially continuous reaction-diffusion equation.

\section{Spectrum of the coupling matrix}
The spatial patterns that can take place on fractal growth
networks are determined by the eigenmodes of the coupling matrix
${\bf M}$, similarly to reaction-diffusion processes on regular
Euclidean lattices \cite{WK1984,K85}. On the other hand, the
stability of the synchronized states is related to the set of
eigenvalues of ${\bf M}$.

In order to analyze the eigenvector problem, consider a fractal
growth network embedded in an Euclidean space of dimension $d$, at
level of construction $L$, on which a spatiotemporal dynamics has
been defined in the vector form of Eq.(\ref{eq:vector}). The
complete set of orthonormal eigenvectors of the corresponding
matrix ${\bf M}$ can be described as the superposition of two
distinct subsets of eigenmodes. One subset, which will be denoted
by $\{{\bf u}_{m n}\}$, contains those eigenvectors associated to
non-degenerate eigenvalues; and the other subset comprises the
eigenvectors corresponding to degenerate eigenvalues of ${\bf M}$,
and will be represented by $\{{\bf v}_{m n}^g\}$ (the indices
refer to the degeneracy. as explained bellow). Thus, the complete
set of eigenvectors of ${\bf M}$ is $\{{\bf u}_{m n}\} \cup \{{\bf
v}_{m s}^g\}$. Each eigenvector describes a basic spatial pattern
that may arise on a fractal growth network characterized by an
embedding dimension $d$ and a level of construction $L$. As
illustration, we shall show the eigenvalues and eigenvectors of
the coupling matrix corresponding to the fractal growth network
embedded in the plane.

\subsection*{Non-degenerate eigenvectors}
The eigenvectors of ${\bf M}$ belonging to the non-degenerate
subset $\{{\bf u}_{m n}\}$ satisfy
\begin{equation}\label{ev1}
 {\bf M} {\bf u}_{m n} = b_{m n} {\bf u}_{mn}, \quad
 \left\{
 \begin{array}{ll}
   m=0; \, n=0  & \\
   m=1,\ldots,L; & n=1,\ldots,3^{m-1};\\
 \end{array}
 \right.
\end{equation}
where $b_{m n}$ is a eigenvalue associated to the eigenvector
${\bf u}_{m n}$.  The $i$ component of vector ${\bf u}_{m n}$ is
$[{\bf u}_{m n}]_i= {\bf u}_{m n}(\alpha_1\alpha_2\dots\alpha_L)$,
according to the rule in Eq.~(\ref{eq:rule}). The number of
eigenvectors in this subset and that of their corresponding
eigenvalues, denoted by $\nu(b)$, is
\begin{equation} \label{eq:numero_bmn}
\nu(b) = 1+\sum\limits_{m=1}^L 3^{m-1} = \frac{3^L + 1}{2} \;.
\end{equation}
Any eigenvector ${\bf u}_{m n}$ in the non-degenerate subset
$\{{\bf u}_{m n}\}$ is characterized by the following property:
its components corresponding to cells of the same type on the
network and located at the same distance (measured in number of
cells) from the center cell $(0^L)$ are identical. Because of this
property, we also refer to the elements in the subset $\{{\bf
u}_{m n}\}$ as {\it symmetric} eigenvectors.

The index $m$ corresponds to the level of construction at which
the eigenvalue $b_{m n}$ first appears, and it is related to the
length of the longest diagonals that arise as the fractal growths
up to the level $L$. At a level of construction $L>1$ there is a
total of $(2^d+1)^{L-1}+1$ diagonals on the network, which can be
classified in $L$ distinct families or groups according to their
lengths, measured in terms of number of cells on the diagonal.
Each different family of diagonals can be identified by the index
$m=1,2,\ldots,L$, and each family contains all the diagonals of
the network that have $3^m$ cells, and thus can support a
wavelength of $3^m$. The index $m$ gives the step in the
construction process at which the family of diagonals of length
$3^m$ has first appeared. For a network defined at level $L$,
there are $(2^d\times (2^d+1)^{L-m-1})$ diagonals in each family
characterized by $m=1,2\ldots,L-1$, and $2$ diagonals in the
family corresponding to $m=L$. For example, in Fig.~1(a) with
$d=2$ and $L=3$, there are $3$ families of diagonals according to
their lengths: the family $m=1$ contains $20$ short diagonals with
a length of $3$ cells; the family $m=2$ has $4$ medium diagonals
measuring $9$ cells; and the family $m=3$ possesses $2$ long
diagonals with $27$ cells. Additionally, we must count the family
$m=0$ associated to diagonals having $1$ element; this corresponds
to the spatially homogeneous eigenvector. Thus, the index
$m=0,1,\ldots,L$, indicates the level of construction at which the
family of diagonals characterized by the index $m$ and having the
same length of $3^m$ cells has appeared. The family of diagonals
identified by $m$ remains in the fractal as the network grows up
to level $L$.

The index $n$ counts the number of distinct symmetric eigenvectors
that have originated each time a new family of diagonals of length
$3^m$ cells appear, and it depends on the number of intersections
that a new diagonal of length $3^m$ has with the other diagonals
already present in the network. There is one intersection for each
$3$ cells in a diagonal; thus the number of intersections is
$3^{m-1}$. These intersections determine the different wavelengths
that can be formed on a diagonal of length $3^m$. The number of
these different wavelengths having reflection symmetry about the
long diagonal of length $3^m$ originated at step $m$ can be
counted by $n=1,2,\ldots,3^{m-1}$.

The spatially homogeneous eigenvector of the matrix ${\bf M}$,
which we denote by ${\bf u}_{0 0}$, belongs to the subset $\{{\bf
u}_{m n}\}$ and its components $N$ are
\begin{equation}
{\bf u}_{00}(\alpha_1\dots\alpha_L) = N^{-1/2}; \qquad \forall L,
     \quad \forall \alpha_k \,
\label{snowflake_homoeigenvectorcomp}
\end{equation}
and
\begin{equation}\label{homo}
    {\bf u}_{00}= \frac{1}{\sqrt{N}} \mbox{col}(1,1,\ldots,1).
\end{equation}
Since the eigenvectors of ${\bf M}$ are mutually orthogonal, all
others eigenmodes in either subset $\{{\bf u}_{mn}\}$ or $\{{\bf
v}_{mn}^g\}$ must satisfy
\begin{equation}
\begin{array}{ll}
    \sum\limits_{\alpha_1,\dots,\alpha_L}
    {\bf u}_{mn}(\alpha_1\dots\alpha_L) = 0; &
    \forall \alpha_k, \quad m \neq 0;
\cr \\
    \sum\limits_{\alpha_1,\dots,\alpha_L}
    {\bf v}_{mn}^g(\alpha_1\dots\alpha_L) = 0; & \forall \alpha_k;
\end{array}
\label{eq:orthogonality}
\end{equation}
that is, the sum over the components of any eigenvector of ${\bf
M}$, different of ${\bf u}_{00}$, is zero.

Figure~2(a) shows the $\nu(b)=5$ non-degenerate eigenvectors and
their associated eigenvalues of the coupling matrix corresponding
to a fractal growth network embedded in dimension $d=2$ and having
a level of construction $L=2$.
Figure~2(b) shows the subset of degenerate eigenvectors ${\bf
v}_{m s}^g$ and the eigenvalues corresponding to a fractal growth
network embedded in an Euclidean space of dimension $d=2$, at
level of construction $L=2$.
\begin{figure}[ht]
\centerline{\hbox{
\epsfig{file=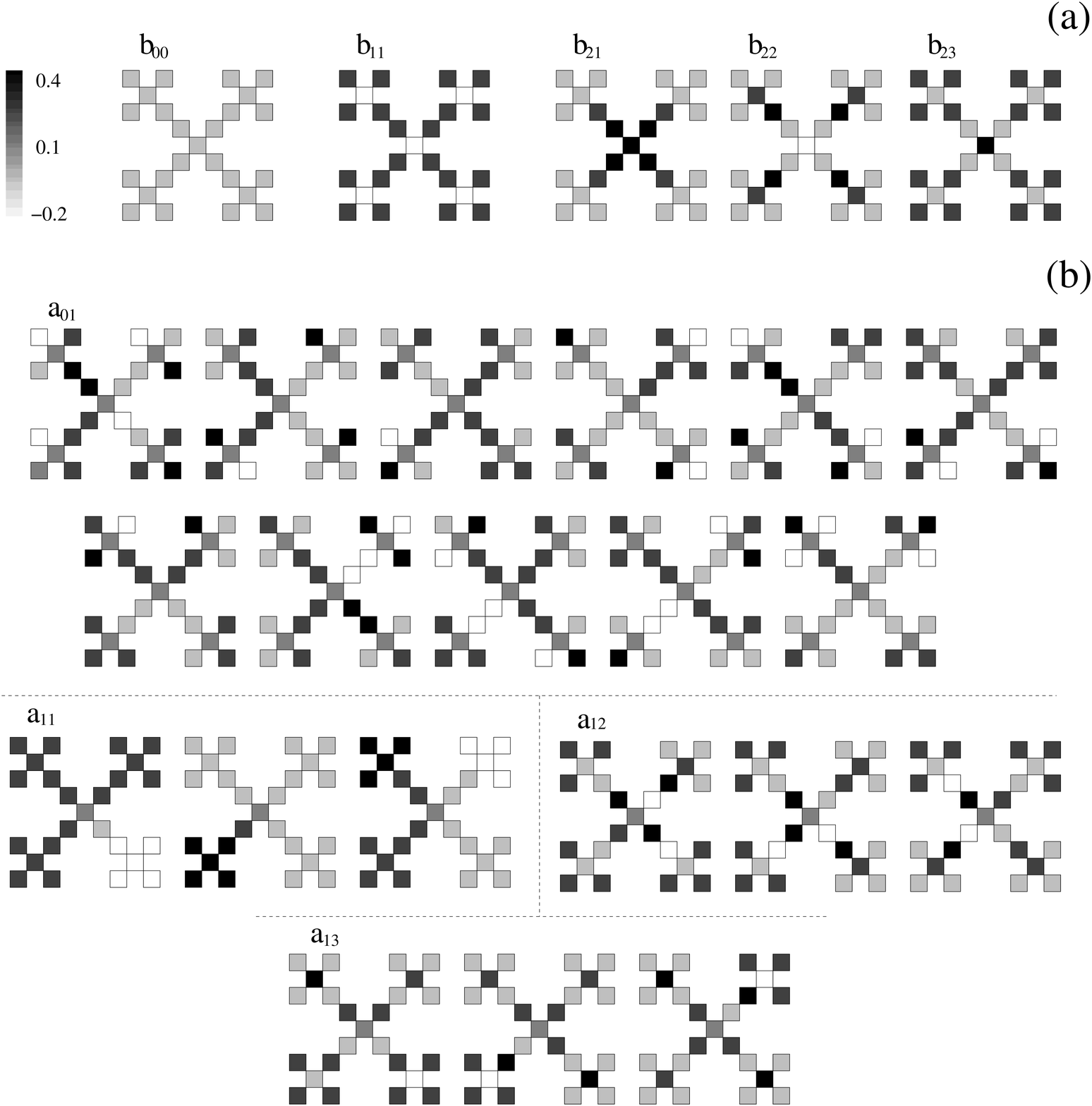, width=0.8\linewidth, angle=0, clip=}}}
\caption{Fractal growth network corresponding to embedding
dimension $d=2$ and level of construction $L=2$. (a) The five
non-degenerated, symmetric eigenvectors ${\bf u}_{m n}$ and their
corresponding eigenvalues. (b) Degenerate eigenvectors ${\bf v}_{m
n}$ and their corresponding eigenvalues.}
\end{figure}

The set of eigenvalues arising from Eq.~(\ref{ev1}) may be ordered
by decreasing value. By Gershgorin's theorem~\cite{B1990}, the
homogeneous eigenvector possesses the largest eigenvalue of ${\bf
M}$, which is $b_{0 0}=0$. In the case of embedding dimension
$d=2$, the smallest eigenvalue for large $L$ is found to be
\begin{equation}
\lim_{L\rightarrow\infty} b_{L\,3^{L-1}} = - (4 + \sqrt{2})\, .
\end{equation}

For each index $m>1$, the eigenvalues $\{b_{m1},b_{m2},\ldots,b_{m
\, 3^{m-1}}\}$ arise in groups of three, $b_{mn}$,  $b_{mn'}$ y
$b_{mn''}$, where
\begin{equation}
n'  = 2\times 3^{m-1}-n+1; \quad \quad n'' = 2\times 3^{m-1}+n.
\end{equation}
The sum of the eigenvalues in each of these groups is constant,
and gives
\begin{equation}
\label{eq:relacion_bmn}
 b_{mn} + b_{mn'} + b_{mn''}= -(2^d + 4) .
\end{equation}
Because of (\ref{eq:relacion_bmn}), the eigenvalues associated to
the non-degenerate eigenvectors satisfy
\begin{eqnarray}
\sum_{m=0}^L \sum_{n=0}^{3^{m-1}} b_{mn} && 
    = -(2^d + 4)\sum_{m=1}^L 3^{m-1} \nonumber \\
 && = -(2^{d-1} + 2)(3^L-1)\, .
\end{eqnarray}

Figure~3 shows the spectrum of eigenvalues $\{ b_{m n}\}$,
indicated by empty symbols, for a fractal growth network embedded
in an Euclidean space of dimension $d=2$, at successive
construction levels $L$. Eigenvalues associated to degenerate
eigenvectors of the coupling matrix ${\bf M}$, to be discussed
next, are also shown in Fig.~3.
\begin{figure}[ht]
\centerline{\hbox{
\psfig{file=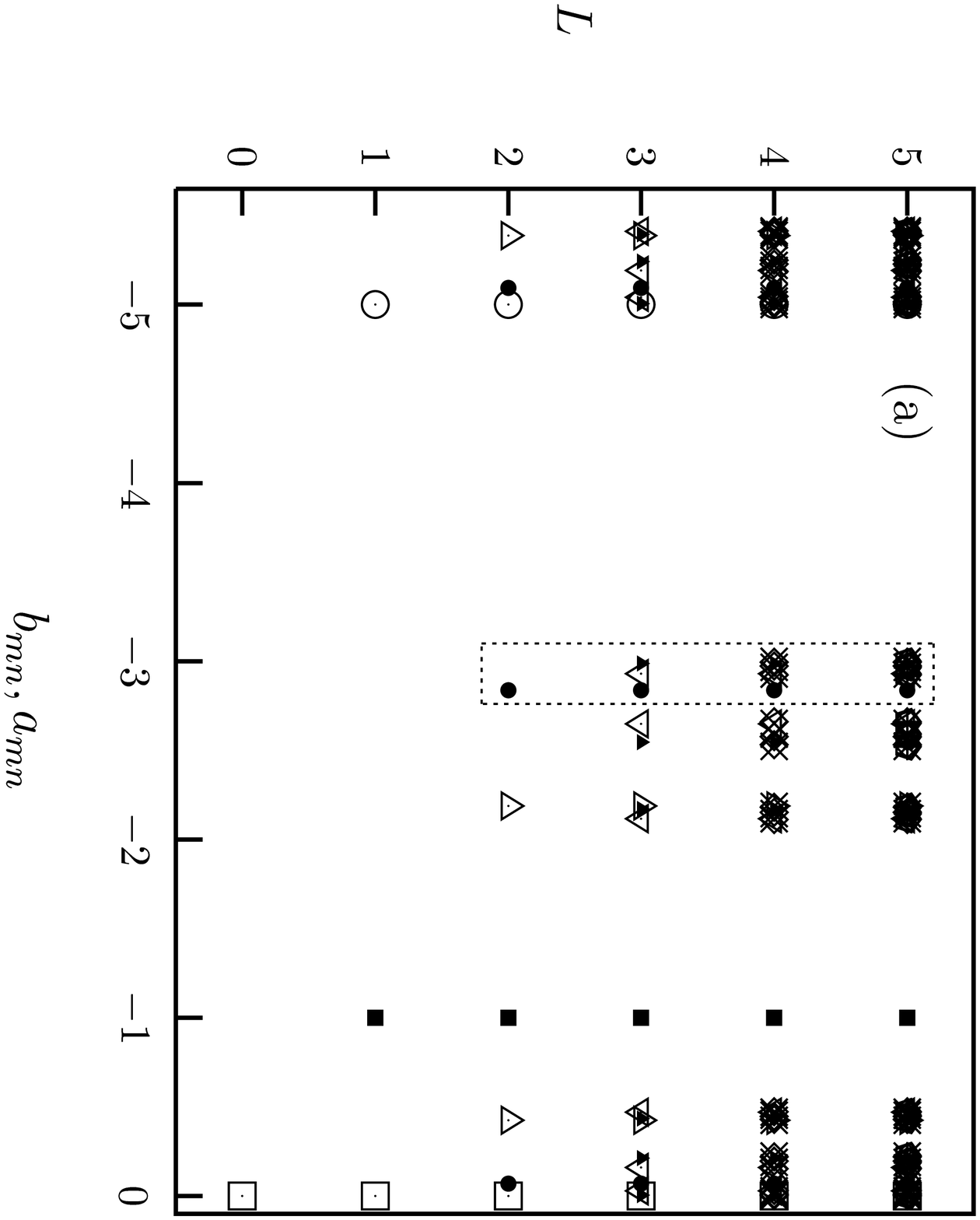,width=0.7\linewidth, angle=90, clip=}}}
\centerline{\hbox{
\psfig{file=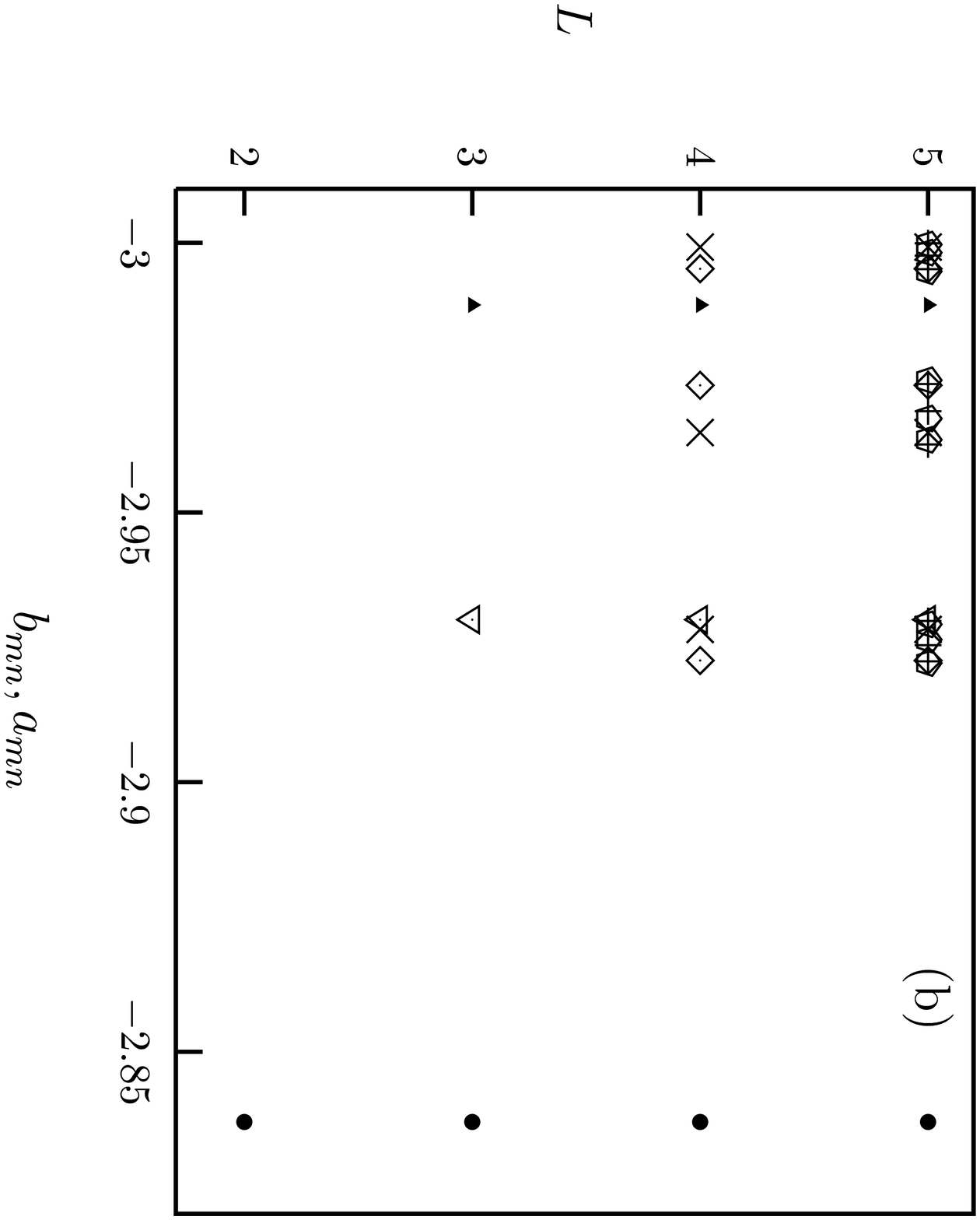,width=0.7\linewidth, angle=90, clip=}}}
\caption{Spectrum of eigenvalues of the coupling matrix at
increasing construction levels $L$, for a fractal growth network
embedded in dimension $d=2$. (a) Eigenvalues $b_{m n}$ are
indicated by empty symbols: $b_{0 0}$ $(\boxdot)$; $b_{1 1}$
$(\odot)$; $b_{2 n}$ $(\triangle)$; $b_{3 n}$ $(\nabla)$; $b_{4
n}$ $(\Diamond)$. Other symbols indicate eigenvalues $a_{m n}$ as
follows: $a_{1 n}$ $(\blacksquare)$; $a_{2 n}$ $(\bullet)$; $a_{3
n}$ $(\blacktriangle)$; $a_{4 n}$ $(\times)$; $a_{5 n}$ $(+)$. (b)
Magnification of the dotted region in (a).}
\end{figure}

\subsection*{Degenerated eigenvectors}
The subset of degenerate eigenvectors $\{{\bf v}_{mn}^g\}$ of the
matrix ${\bf M}$ satisfy
\begin{equation}
\begin{array}{l}
{\bf M}{\bf v}_{mn}^g = a_{mn} {\bf v}_{mn}^g , \\
   m=1,2,\ldots,L; \quad n=1,2,\ldots,3^{m-1};
   \end{array}
\label{eq:deg_autovec_def}
\end{equation}
where $a_{m n}$ is the eigenvalue associated to a group of
$\Omega$ degenerate eigenvectors $\{{\bf v}_{m n}^1, {\bf v}_{m
n}^2, \dots, {\bf v}_{m n}^\Omega\}$ belonging to the subset
$\{{\bf v}_{m n}^g\}$. The index $g$ runs from $1$ to a value
$\Omega$ and counts the different eigenvectors associated to the
degenerate eigenvalue $a_{m n}$, as it will be shown bellow. The
integer indices $m$ and $n$ label different eigenvalues $a_{m n}$.

The $i-$component of a vector ${\bf v}_{m n}^g$ corresponds to a
cell of the network labeled by the rule Eq.~(\ref{eq:rule}), i.e.,
$[{\bf v}_{m n}^g]_i = {\bf v}_{m n}^g(\alpha_1 \alpha_2 \dots
\alpha_L)$. The eigenmodes in the subset $\{{\bf v}_{m n}^g\}$ are
characterized by the following two properties,
\begin{equation}
{\bf v}_{mn}^g (\alpha_1 \dots \alpha_{L-m}0^m)=0 ;  \quad
m=1,2,\dots,L-1 ; \label{eq:asim_center_prop}
\end{equation}
that is, all the components of ${\bf v}_{m n}^g$ corresponding to
center cells formed in the first $L-m$ levels of construction
vanish; and
\begin{equation}
\sum_{\alpha_1, \alpha_2, \dots, \alpha_{L-1}}
       {\bf v}_{mn}^g (\alpha_1 \alpha_2 \dots \alpha_{L-1} 0) = 0 \,;
\label{eq:asim_sum_property}
\end{equation}
that is, the sum of the components associated to the center cells
in a network spatially described by a vector ${\bf v}_{m s}^g$, is
zero.

In contrast to the eigenvectors in the non-degenerate subset, an
eigenvector ${\bf v}_{m s}^g$ is non-symmetric; however it
exhibits a partial regularity, having all its components
corresponding to center cells equal to zero if they originated at
the level of construction $L-m$. The index $m$ counts the number
of additional levels of construction in which non-vanishing center
cells have appear up to level $L$; and its possible values are
$m=1,2,\ldots,L$. There are $(2^d+1)^{L-m}$ vanishing center cells
in an eigenvector ${\bf v}_{m s}^g$. Each null center cell in the
eigenvector is the center of $2^{d-1}-1$ diagonals, except the
cell $(0^L)$ that is the center of one additional diagonal. Thus,
adding the additional diagonal of the center cell $(0^L)$, there
are $(2^d+1)^{L-m}(2^{d-1}-1)+1$ diagonals whose centers are null
center cells. The index $n$ counts the number of possible
intersections that these diagonals have with other diagonals; its
possible values are $n=1,2,\ldots,3^{m-1}$.

Because an eigenvector ${\bf v}_{m s}^g$ is a non-symmetric mode,
the $2$ edge cells at the ends of each of the
$(2^d+1)^{L-m}(2^{d-1}-1)+1$ diagonals having null center cells
are independent of each other. Therefore, there exist
$2[(2^d+1)^{L-m}(2^{d-1}-1)+1]$ edge cells with this
characteristics, for given values of $m$ and $n$. However, due to
the orthogonality property, Eq.~(\ref{eq:orthogonality}), the
total number of linearly independent eigenvectors ${\bf v}_{m
s}^g$ possessing the same indices $m$ and $n$ is
$\Omega=2[(2^d+1)^{L-m}(2^{d-1}-1)+1]-1$. The index $g$ counts the
number of independent eigenvectors associated to an eigenvalue
$a_{m n}$, and it may take the values $g=1,2,\ldots,\Omega$. In
this fashion, the subset of degenerate eigenvectors $\{{\bf v}_{m
s}^g\}$ of the coupling matrix ${\bf M}$ corresponding to a
fractal growth network embedded in dimension $d$ and at level of
construction $L$ is fully described.

In the case of embedding dimension $d=2$, there are $5^{L-m}$
vanishing center cells in an eigenvector ${\bf v}_{m s}^g$ and
there are $5^{L-m}+1$ diagonals whose centers are null center
cells. The number of linearly independent eigenvectors ${\bf v}_{m
s}^g$ having the same indices $m$ and $n$ is
$\Omega=2(5^{L-m}+1)-1= 2 \times 5^{L-m} +1$. Thus, the index $g$
may take the values $g=,2,\ldots,2 \times 5^{L-m} +1$.

The number of different eigenvalues, $a_{mn}$, that belong to the
spectrum of matrix ${\bf M}$ for a fractal growth network at
construction level $L$, is
\begin{equation}
\nu(a)=\sum_{m=1}^L 3^{m-1} = \frac{3^L-1}{2}\;.
 \label{eq:numero_amn}
\end{equation}
At each level $m>1$, there appear $3^{m-1}$ new eigenvalues
$a_{mn}$. Similarly to the non-degenerate eigenvalues, the $a_{m
n}$ can be grouped in sets of three eigenvalues that satisfy
\begin{equation}
\label{eq:relacion_amn1}
 a_{mn} + a_{mn'} + a_{mn''}= -(2^d + 4),
\end{equation}
where
\begin{equation}
\label{eq:relacion_amn2}
  n'  = 2\times 3^{m-1}-n+1 \, \quad
  n'' = 2\times 3^{m-1}+n \, ;
\end{equation}
and therefore the total sum of the degenerate eigenvalues $a_{mn}$
of a matrix ${\bf M}$ corresponding to a fractal growth network
embedded in a dimension $d$ and at level of construction $L$ gives
\begin{eqnarray}
\sum_{m=1}^L \sum_{n=1}^{3^{m-1}} a_{mn} && 
  = -(2^d + 4) \,\sum_{m=1}^L 3^{m-1}    \nonumber \\ 
  && = -(2^{d-1} + 2)(3^{L-1}-1)\,.
\end{eqnarray}

Figure~3 shows the spectrum of eigenvalues $\{a_{m n}\}\cup\{b_{m
n}\}$ for a fractal growth network embedded in Euclidean space of
dimension $d=2$ at successive levels of construction $L$. The
distribution of the spectrum of eigenvalues of the coupling matrix
${\bf M}$ can be seen as a function of the growth process of the
fractal structure. Note that the full spectrum $\{a_{m
n}\}\cup\{b_{m n}\}$ is always contained between the eigenvalues
$b_{00}=0$ and $b_{L \, 3^{L-1}}$.

From Eqs.~(\ref{eq:numero_bmn}) and (\ref{eq:numero_amn}), the
total number of distinct eigenvalues of ${\bf M}$, including both
types $a_{mn}$ and $b_{mn}$, and denoted by $\nu({\bf M})$ is
given by
\begin{equation}
 \label{eq:snowflake_num_eigenvalues}
\nu({\bf M})= \nu(b)+\nu(a)= \frac{3^L + 1}{2} + \frac{3^L-1}{2} =
3^L.
\end{equation}
Since there appear $3^{m-1}$ new eigenvalues of type $a_{m n}$ at
each level of construction $m$, and there are
$\Omega=2[(2^d+1)^{L-m}(2^{d-1}-1)+1]-1$ eigenvectors ${\bf v}_{m
s}^g$ associated to each eigenvalue $a_{m n}$, the total number of
independent eigenvectors in the subset $\{{\bf v}_{mn}^g\}$ is
\begin{equation}
\sum_{m=1}^L \Omega \left( 3^{m-1} \right) = (2^d+1)^L -
\frac{3^L+1}{2}\,.
\end{equation}

From Eq.~(\ref{eq:numero_bmn}) we know that the number of
independent eigenvectors in the subset $\{{\bf u}_{mn}\}$ is
$(3^L+1)/2$. Therefore, the total number of independent
eigenvectors of ${\bf M}$ is
\begin{equation}
\frac{3^L+1}{2} + (2^d+1)^L - \frac{3^L+1}{2} =(2^d+1)^L =N,
\end{equation}
as should be expected.

Figure~4(a) shows the complete spectrum of eigenvalues of ${\bf
M}$ and the degeneracy fraction of each eigenvalue, for a fractal
growth network embedded in an Euclidean space of dimension $d=2$,
at level of construction $L=5$. The degeneracy $\Omega= 2 \times
5^{5-m} +1$ of each of the $\nu(a)=(3^5-1)/2=121$ eigenvalues
$a_{mn}$ is plotted as a vertical bar, while the
$\nu(b)=(3^5+1)/2=122$ different eigenvalues $b_{m n}$ are
indicated by plus symbols $(+)$ and they are non-degenerate. It is clear that
both the distribution of eigenvalues and their degeneracies are
nonuniform. The scaling properties of the spectrum of eigenvalues
of the coupling  matrix can also be conveniently represented by
plotting the accumulated sum of the degeneracies of all
eigenvalues, that is, the measure of the spectrum of ${\bf M}$
(denoted by $\rho$), on the eigenvalue axis for large $L$, as in
Fig.~4(b). The resulting graph exhibits the features of a devil
staircase, a fractal curve arising in a variety of nonlinear
phenomena.
\begin{figure}[ht]
\centerline{\hbox{
\psfig{file=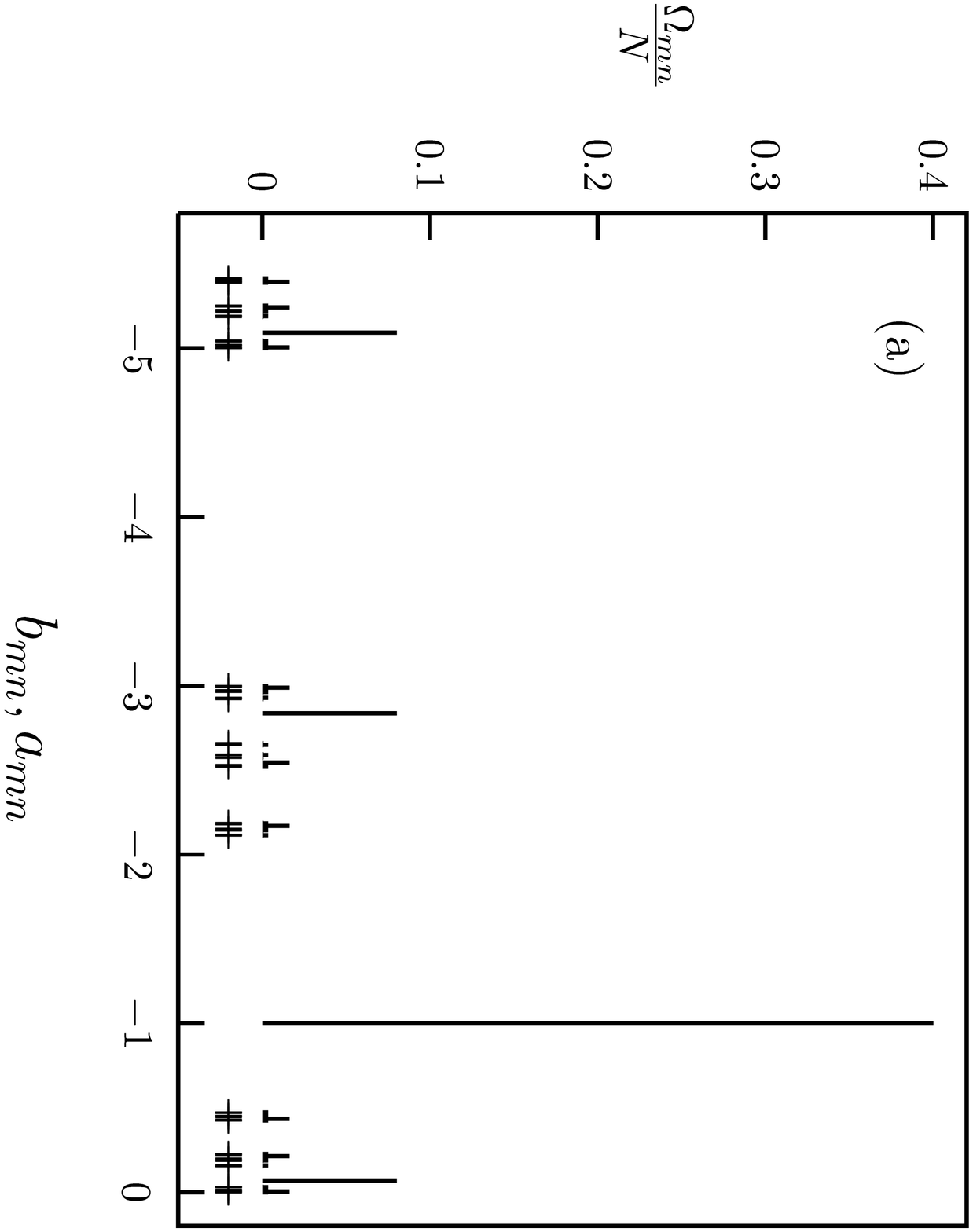, width=0.70\linewidth, angle=90, clip=}}}
\centerline{\hbox{
\psfig{file=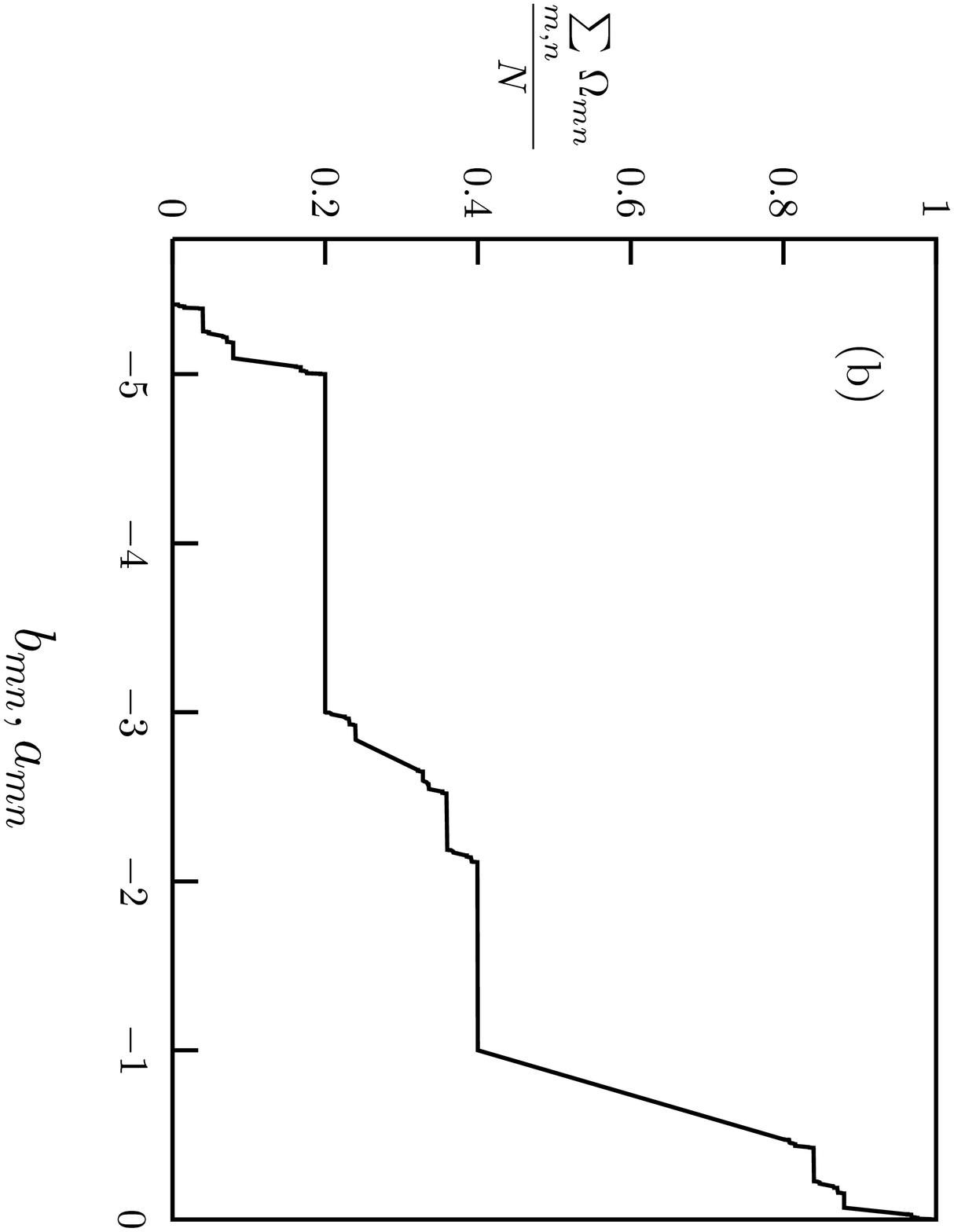, width=0.70\linewidth, angle=90, clip=} }}
\caption{(a) Distribution and degeneracies of the spectrum of
eigenvalues of the coupling matrix ${\bf M}$ for a fractal growth
network embedded in $d=2$ at level of construction $L=5$. The
eigenvalues $b_{m n}$ are indicated with $(+)$ just below the zero
line, for clarity. The vertical axis shows the degeneracy of the
eigenvalues $\{a_{mn}\}$ divided by $N$, indicated by a vertical
bar at each eigenvalue. (b) The measure of the full set of
eigenvalues of ${\bf M}$.}
\end{figure}

The eigenvectors of the coupling matrix reflect the topology of
the fractal growth network and they are analogous to the Fourier
eigenmodes appearing in regular Euclidean lattices. In this sense,
the symmetry properties of  the non-degenerate eigenvectors
$\{{\bf u}_{ m n}\}$ and the conditions given in
Eqs.~(\ref{eq:asim_center_prop})-(\ref{eq:asim_sum_property}) for
the degenerate eigenvectors  $\{{\bf v}_{m s}^g\}$ represent
different wavelengths on a fractal growth network embedded in an
Euclidean space of dimension $d$ at level of construction $L$.

\section{Synchronized states and pattern selection}
Synchronized states in spatiotemporal systems are relevant since
we are often interested in mechanisms by which a spatial pattern
can be selected in a uniform system that breaks its symmetry as a
parameter is changed.

Consider spatially synchronized, period-$K$ states such as
$x_t(\alpha_1 \ldots\alpha_L)= \bar{x}_k$, $\forall (\alpha_1
\ldots\alpha_l)$; where $\bar{x}_k$, $(k=1,2,\ldots,K)$, is a
period-$K$ orbit of the the local map, satisfying
$f^{(K)}(\bar{x}_k)=\bar{x}_k$. The linear stability analysis of
periodic, synchronized states in coupled map lattices is carried
out by the diagonalization of  the matrix ${\bf M}$ in
Eq.~(\ref{eq:vector}), and it leads to the conditions
\cite{WK1984}
\begin{equation}
\label{eq:stability} \prod_{k=1}^K \left| f'(\bar{x}_k) + \gamma
\mu \right| < 1\;,
\end{equation}
where $\mu$ is any of the $\nu({\bf M})=3^L$ eigenvalues,
associated to either subset of eigenvectors $\{{\bf u}_{m n}\}$ or
$\{{\bf v}_{m s}^g\}$, of the coupling matrix ${\bf M}$
corresponding to a fractal growth network embedded in a
$d$-dimensional Euclidean space at level of construction $L$.

The nonuniform distribution of the eigenvalue spectrum is
manifested in the stability of the synchronized states through
Eq.~(\ref{eq:stability}) and gives rise to marked differences when
compared, for instance, with the bifurcation structure on regular
lattices. As an application, consider a local dynamics described
by the logistic map, $f(x)=\lambda x (1-x)$. In this case, the
stability conditions, Eq.~(\ref{eq:stability}), for the period
$K=2^p$, synchronized state yield the set of boundary curves
\begin{equation}
\label{eq:synlog}
 S_L^p(\mu) \equiv \prod_{k=1}^{2^p}
    \left[ \lambda(1-2\bar{x}_k) + \gamma \mu \right]= \pm 1\;.
\end{equation}
For each sign, Eqs.~(\ref{eq:synlog}) give $3^L$ boundary curves
in the plane $(\gamma,\lambda)$, corresponding to the different
values of $\mu$; these curves determine the stability regions of
the period-$2^p$, synchronized states on the network.

The scaling structure for the period-$2^p$, synchronized states in
fractal growth networks is similar to that of a any lattice
described by a diffusive coupling matrix, since the form of
Eq.~(\ref{eq:synlog}) is the same in any case. As for regular
Euclidean lattices \cite{WK1984,K85}, the stability regions for
the period-$2^p$, synchronized states in the $(\gamma, \lambda)$
plane scale as $\lambda \sim \delta^{-p}$, and $\gamma \sim
\alpha^{-p}$, where $\delta=4.669\ldots$ and $\alpha=-2.502\ldots$
are Feigenbaum's scaling constants for the period doubling
transition to chaos. However, the specific structure of the
eigenvalue spectrum of the coupling matrix determines the shapes
and gaps of the regions of stability of synchronized, periodic
states.

From Eqs.~(\ref{eq:synlog}), the boundary curves for the
synchronized, fixed point state $(p=0)$ on the parameter plane
$(\gamma,\lambda)$ are given by the straight lines
\begin{equation}
\lambda=\mu \gamma +1\;, \quad\quad\quad \lambda=\mu \gamma +3\;;
\end{equation}
which are first crossed for the most negative eigenvalue,
$\mu=b_{L 3^{L-1}}$. Similarly, the boundaries period-two ($p=1$),
synchronized state on a fractal growth network embedded in an
Euclidean space of dimension $d=2$ and at construction level $L=3$
are given by the two sets
\begin{equation}\label{eq:boundary1}
S_3^1(a_{mn})= -\lambda^2+2\lambda+4+
                      \gamma a_{mn}(\gamma a_{mn}-2)=
  \pm 1 \, ,
\end{equation}
and
\begin{equation}\label{eq:boundary2}
 S_3^1(b_{mn})= -\lambda^2+2\lambda+4+\gamma b_{mn}(\gamma
b_{mn}-2)= \pm 1 \,;.
\end{equation}

Figure~5(a) shows the boundary curves
Eqs.~(\ref{eq:boundary1})-(\ref{eq:boundary2}) on the plane
$(\gamma, \lambda)$. The boundary between the synchronized, fixed
point state and the synchronized period-two state occurs at
$\lambda=3$. The upper boundaries (corresponding to $-1$ in the
r.h.s of Eqs.~(\ref{eq:boundary1})-(\ref{eq:boundary2})) have
minima $\lambda_{\min}=1+\sqrt{5}$ at values $\gamma_{\min}=1/b_{m
n}$ and $\gamma_{\min}=1/a_{m n}$ (for any period $2^p$,
$\lambda_{\min}$ depends on $p$). Figure~5(b) shows a
magnification of Fig.~5(a) around the minima of the upper
boundaries. The distribution of the minima $\gamma_{\min}$ and the
presence of nonuniformly distributed gaps in the boundary curves
is a manifestation of the nonuniform structure of the eigenvalue
spectrum. Since the nonuniformity in the distribution of
eigenvalues persists at any construction level $L$ of a fractal
growth network, this property allows for regions of stability of
the synchronized states or gaps characteristic of fractal networks
and which are not present in other geometries, for example in
regular lattices, where the distribution of eigenvalues of the
coupling matrix is uniform and continuous in the limit of infinite
size lattices.

The set of eigenvectors $\{{\bf u}_{m n}\} \cup \{{\bf v}_{m
n}^g\}$ of the coupling matrix ${\bf M}$ constitute a complete
basis (normal modes) on which a state ${\bf x}_t$ of the system
can be represented as a linear combination of these vectors. Thus.
the evolution of ${\bf x}_t$ reflects the stabilities of the
normal modes. Figure~5(b) shows how a spatially inhomogeneous
pattern may be selected as the synchronized state becomes unstable
through crossing of the upper boundary; the first boundary segment
crossed determines the character of the instability and the
properties of the selected pattern. For example, consider an
initial state consisting of a small perturbation of the
synchronized, period-$2$ state at parameter values just beyond the
boundary segment corresponding to the eigenvalue $a_{3 6}$,
indicated by a cross in Fig.~5(b), where this initial state is
unstable. The inhomogeneous period-$4$ final spatial pattern is
represented in Fig.~6; it corresponds to a linear combination of
the three eigenmodes $\{{\bf v}_{3 6}^g; \, g=1,2,3 \}$ associated
to the degenerate eigenvalue $a_{3 6}$ of the matrix ${\bf M}$
corresponding to the fractal network embedded in dimension $d=2$
at level of construction $L=3$. All other modes are unstable in
this region of parameter space. For any construction level $L$ of
the fractal network, and any period $2^p$, the boundary curve
$S_3^1(a_{3 6})=-1$ separates a gap of the synchronized state from
the stable region for the eigenmodes ${\bf v}_{3 6}^g$
corresponding to $a_{3 6}$. Therefore, a transition between these
two spatial patterns can always be observed in the appropriate
regions of the parameter plane $(\gamma, \lambda)$.
\begin{figure}[ht]
\centerline{\hbox{
\psfig{file=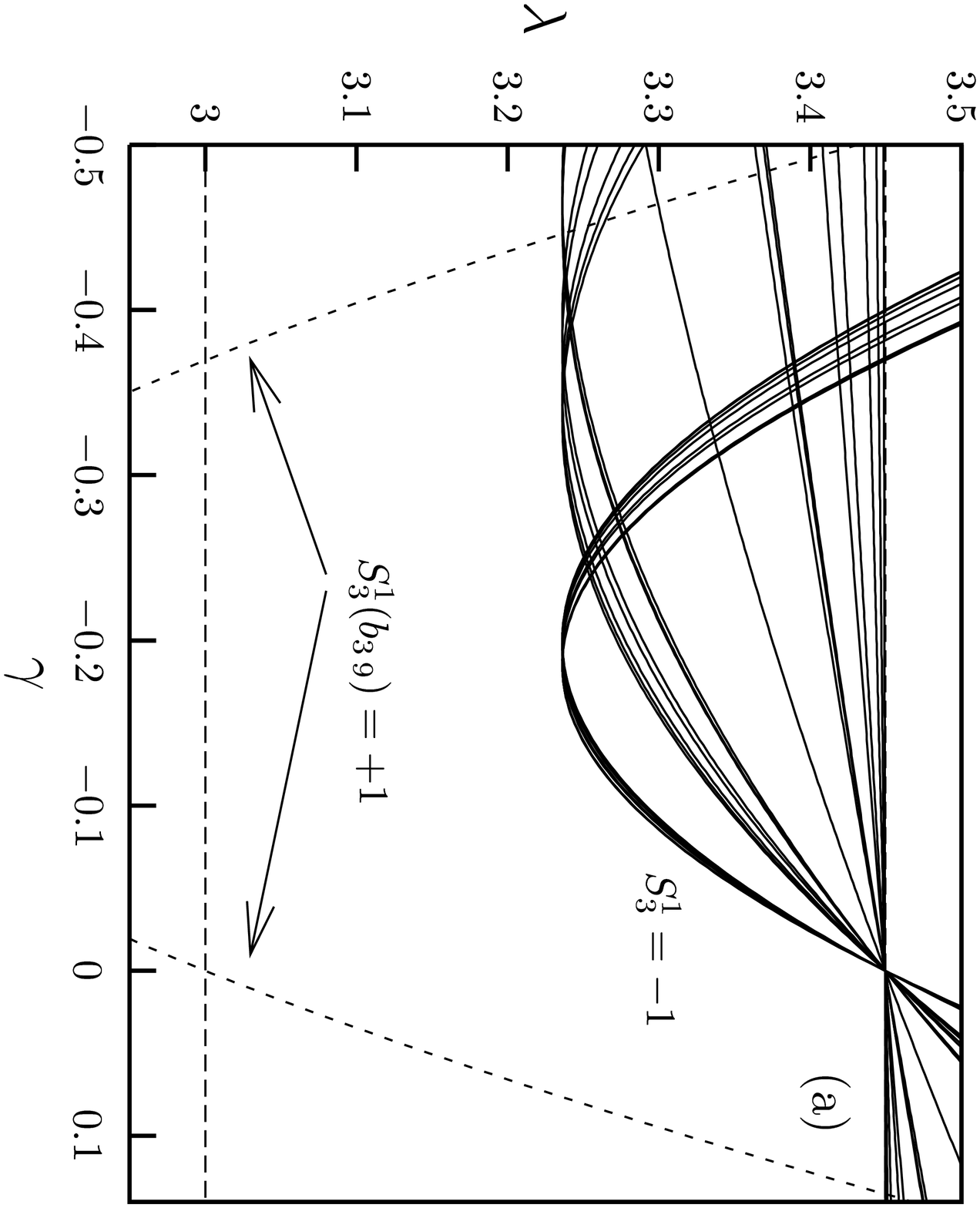, width=0.7\linewidth, angle=90, clip=}}}
\centerline{\hbox{
\psfig{file=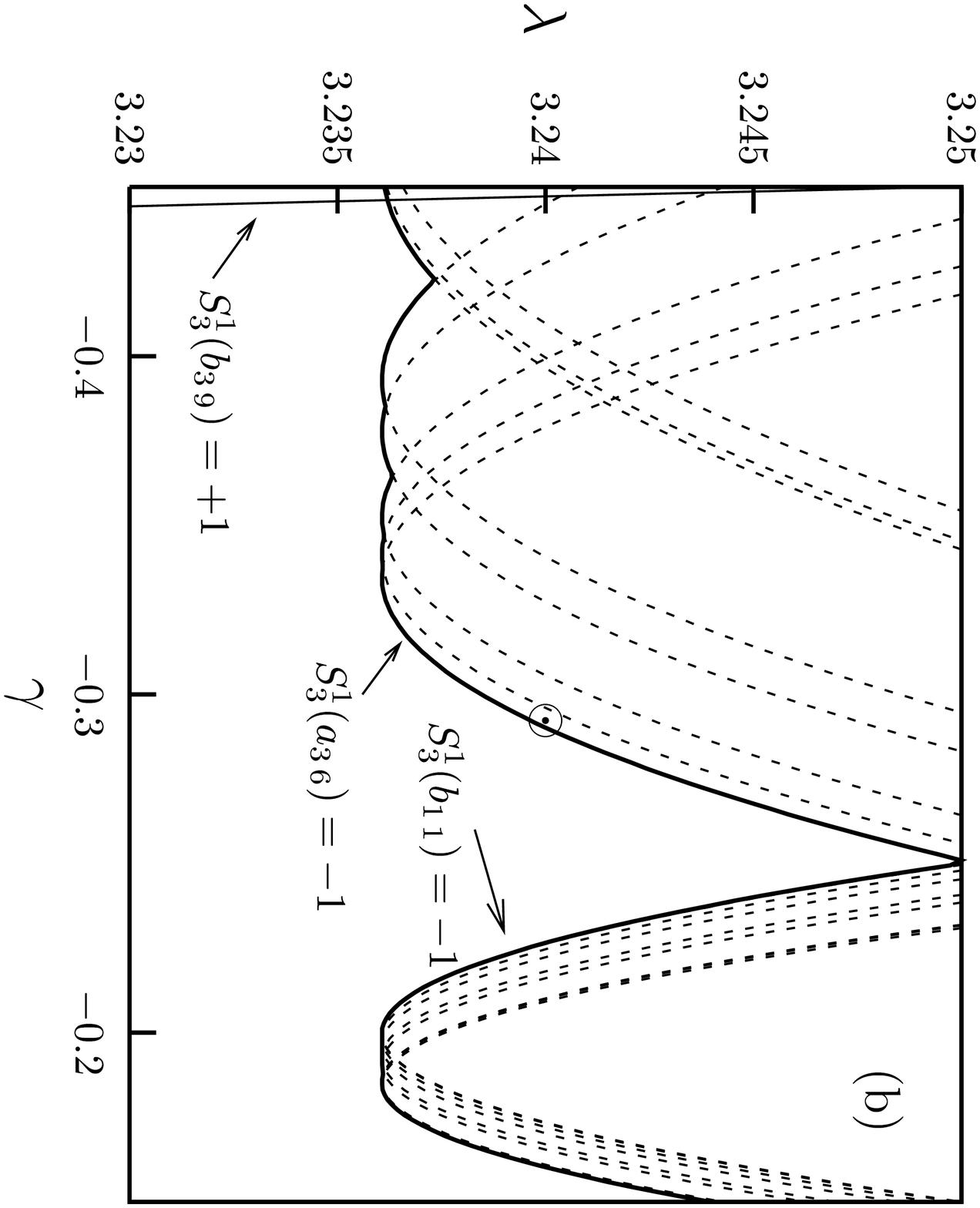, width=0.7\linewidth, angle=90, clip=} }}
\caption{The boundary curves $S_3^1=\pm1$ given by Eqs.
(\ref{eq:boundary1})-(\ref{eq:boundary2}) for the period-two,
synchronized states of a fractal growth network at level of
construction $L=3$, embedded in $d=2$. (a) The upper curves
correspond to the r.h.s of Eqs.
(\ref{eq:boundary1})-(\ref{eq:boundary2}) equal to $-1$ for both
types of eigenvalues. Arrows indicate the boundary curve
$S_3^1(b_{39})=+1$. The interior region bounded by these curves is
where stable, synchronized, period-two states exist in the
parameter plane $(\lambda,\gamma)$. (b) Magnification of the upper
curves in (a),  showing the gaps in the stability boundaries of
the period-two, synchronized states. Arrows indicate boundary
curves corresponding to several eigenvalues. The symbol $(\odot)$
just beyond the boundary $S_3^1(b_{36})=-1$ indicates the values
of the parameters $\gamma$ and $\lambda$ used in Fig.~6.}
\end{figure}

\begin{figure}[ht]
\centerline{\hbox{
\epsfig{file=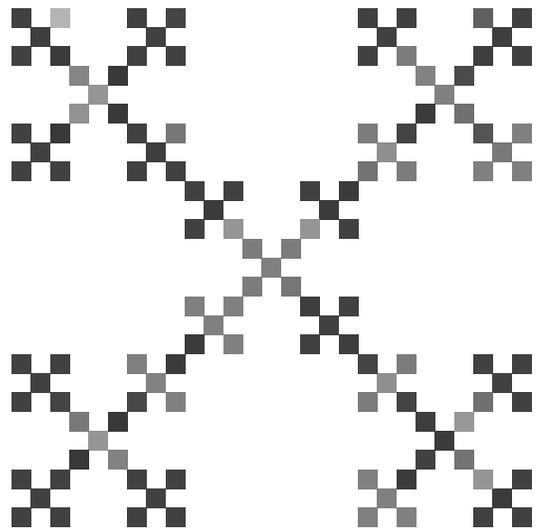, width=0.8\linewidth, angle=0, clip=}}}
\caption{Inhomogeneous, period-$4$ state at parameter values
$\gamma=-0.296$, $\lambda=3.24$. This pattern is a linear
combination of the three eigenvectors associated to the eigenvalue
$a_{3\,6}$.}
\end{figure}

\section{Conclusions}
The underlying inhomogeneous structure of fractal growth networks
has significant effects on the spatial patterns that can be formed
by reaction-diffusion processes on these geometrical supports. In
systems of interacting elements, such as the models considered in
this article, the coupling matrix contains the connectivity of the
network. The set of eigenvectors of the coupling matrix reflect
this connectivity. The spatial patterns that can arise in fractal
growth networks are determined by the eigenvectors of the
diffusion coupling matrix ${\bf M}$, which constitute a complete
basis on which any spatial mode can be expressed as a linear
combination. These eigenvectors have complex spatial forms but
they are analogous to the Fourier eigenmodes arising in regular
Euclidean lattices. On the other hand, the stability of the
synchronized states is determined by the eigenvalues of ${\bf M}$.
The density distribution of these eigenvalues and their degeneracy
are nonuniform. These features affect the bifurcation properties
of dynamical systems such as coupled maps defined on fractal
growth networks. The scaling structure of the synchronized,
period-doubled states on the space of parameters of the system is
similar for both uniform and fractal networks but the nature of
the bifurcation boundaries is different. For fractal growth
networks, the nonuniform distribution of eigenvalues leads to gaps
in the boundary curves separating stable synchronized states that
are not present for coupled maps on uniform lattices, where the
spectrum of eigenvalues is continuous. These gaps allow for the
selection of specific spatial patterns arising from a uniform,
synchronized state by appropriately changing the parameters of the
system.

Although we have presented only the simplest spatiotemporal
patterns that can be formed on fractal growth structures, the
formalism introduced in this article can be applied to many other
processes, such as excitable dynamics, nontrivial collective
behavior, phase-ordering, domain segregation, turbulence, etc. The
formalism is also useful for cellular automata models and
continuous-time local dynamics on fractal growth networks.

The study of dynamical systems defined on nonuniform spatial
supports, such as fractal growth networks, provides insight into
the relationship between topology and spatiotemporal phenomena in
complex networks.

\section*{Acknowledgement}
This work was supported in part by Consejo de Desarrollo
Cient\'{\i}fico, Human\'{\i}stico y Tecnol\'ogico, Universidad de
Los Andes, M\'erida, Venezuela.


\begin{thebibliography}{99}
\bibitem{Kaneko} Chaos 2 (1992) 279, focus issue on
Coupled Map Lattices; edited by K. Kaneko.
\bibitem{CK1} M. G. Cosenza and R. Kapral, Phys, Rev. A  (1992) 1850.
\bibitem{TCA} K. Tucci, M. G. Cosenza and O. Alvarez-Llamoza,
Phys. Rev. E 68 (2003) 027202.
\bibitem{Gade} P. M. Gade, H. A Cerdeira, and R. Ramaswamy, Phys. Rev. E
 52 (1995) 2478.
\bibitem{Tree} M. G. Cosenza and K. Tucci, Phys. Rev. E 64 (2001) 026208.
\bibitem{Vol} D. Volchenkov, S. Sequeira, Ph. Blanchard, and M. G. Cosenza,
Stochastics and Dynamics 2 (2002) 203 .
\bibitem{Kay} M. G. Cosenza and K. Tucci, Phys. Rev. E 65 (2002) 036223.
\bibitem{Amri} S. Jalan and R. E. Amritkar, Phys. Rev. Lett. 90
(2003) 014101.
\bibitem{V1983}  T. Vicsek, J. Phys. A (Math. Gen.) 16 (1983), L647.
\bibitem{Vicsek} T. Vicsek, {\it Fractal Growth Phenomena} (World
Scientific, Singapore, 1989).
\bibitem{WK1984} I. Waller and R. Kapral, Phys. Rev. A 30 (1984)
2047.
\bibitem{K85} R. Kapral, Phys. Rev. A 31 (1985) 3868.
\bibitem{B1990} S. Barnett, {\it Matrices, Methods and
Applications} (Oxford University Press, Oxford, 1990).
\end{thebibliography}
\end{document}